\documentclass[aps,prl,twocolumn,superscriptaddress,showpacs]{revtex4-1}

\usepackage{graphicx}
\usepackage{amsmath}
\usepackage{float}

\begin{document}


\title{Measuring the Transverse Spin Density of Light}


\author{Martin Neugebauer}
\affiliation{Max Planck Institute for the Science of Light, Guenther-Scharowsky-Str. 1, D-91058 Erlangen, Germany}
\affiliation{Institute of Optics, Information and Photonics, University Erlangen-Nuremberg, Staudtstr. 7/B2, D-91058 Erlangen, Germany}
\author{Thomas Bauer}
\affiliation{Max Planck Institute for the Science of Light, Guenther-Scharowsky-Str. 1, D-91058 Erlangen, Germany}
\affiliation{Institute of Optics, Information and Photonics, University Erlangen-Nuremberg, Staudtstr. 7/B2, D-91058 Erlangen, Germany}
\author{Andrea Aiello}
\affiliation{Max Planck Institute for the Science of Light, Guenther-Scharowsky-Str. 1, D-91058 Erlangen, Germany}
\affiliation{Institute of Optics, Information and Photonics, University Erlangen-Nuremberg, Staudtstr. 7/B2, D-91058 Erlangen, Germany}
\author{Peter Banzer}
\email[]{peter.banzer@mpl.mpg.de}
\homepage[]{http://www.mpl.mpg.de/}
\affiliation{Max Planck Institute for the Science of Light, Guenther-Scharowsky-Str. 1, D-91058 Erlangen, Germany}
\affiliation{Institute of Optics, Information and Photonics, University Erlangen-Nuremberg, Staudtstr. 7/B2, D-91058 Erlangen, Germany}
\affiliation{Department of Physics, University of Ottawa, 25 Templeton, Ottawa, Ontario K1N 6N5 Canada}


\date{\today}

\begin{abstract}
We generate tightly focused optical vector beams whose electric fields spin around an axis transverse to the beams' propagation direction. We experimentally investigate these fields by exploiting the directional near-field interference of a dipole-like plasmonic field probe, placed adjacent to a dielectric interface, which depends on the transverse electric spin density of the excitation field. Near- to far-field conversion mediated by the dielectric interface enables us to detect the directionality of the emitted light in the far-field and, therefore, to measure the transverse electric spin density with nanoscopic resolution. Finally, we determine the longitudinal electric component of Belinfante's elusive spin momentum density, a solenoidal field quantity often referred to as 'virtual'.
\end{abstract}


\maketitle

\begin{figure}[htbp]
\centerline{\includegraphics[width=.76\columnwidth]{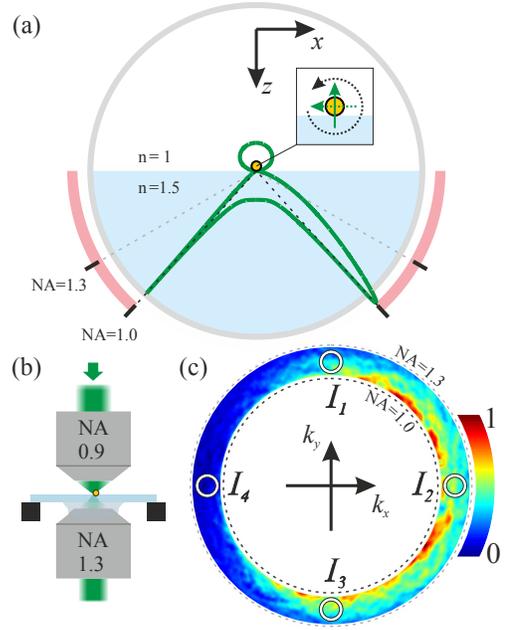}}
\caption{\label{fig_setup}(color). Experimental concept, setup and far-field detection scheme. (a) A transversely spinning dipole (here: counterclockwise spinning denoted by the black arrow in the inset) in close proximity to a dielectric interface has a directional emission pattern (green line) into the region above the critical angle (red arc). The actual critical angle is indicated by the dashed black line at the numerical aperture $\text{NA}=1$. (b) An incoming (here: radially or linearly polarized) paraxial beam is tightly focused onto a gold nanoparticle (radius $=40$ nm) sitting on a glass substrate. The light emitted by the dipole into the region above the critical angle ($\text{NA}>1$) can be collected up to an NA of 1.3 by the immersion-type objective. An exemplarily measured far-field pattern of a transversely spinning dipole is depicted in (c). Only the light scattered by the particle is detected above the critical angle. Averaging the intensity in four small regions in $k$-space (white circles) yields four intensity values ($I_{1}$,$I_{2}$,$I_{3}$, and $I_{4}$).}
\end{figure}

\begin{figure}[htbp]
\centerline{\includegraphics[width=0.8\columnwidth]{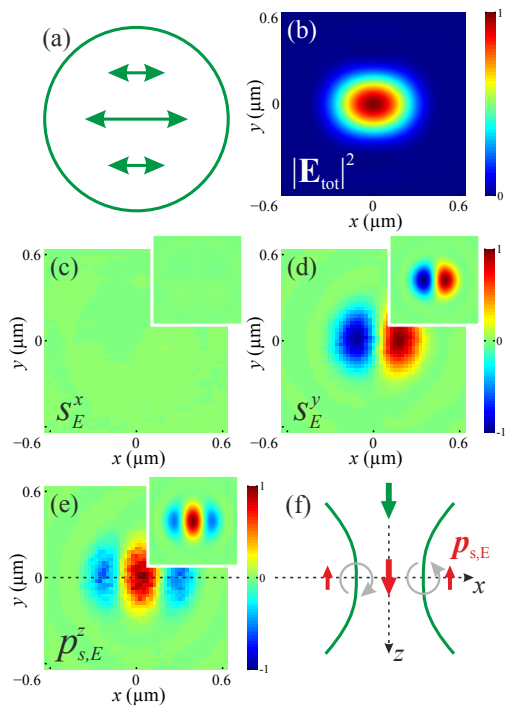}}
\caption{\label{fig_linear}(color). (a) The beam profile of the incoming paraxial linearly polarized beam, and (b) the respective theoretically calculated focal distribution of $\left|\textbf{E}_{\text{tot}}\right|^{2}$. The measured distributions of $s_{E}^{x}$ and $s_{E}^{y}$ (normalized to their common maximum value) are depicted in (c) and (d), and the experimentally reconstructed distribution of $p_{s,E}^{z}$ (normalized to its maximum value) is shown in (e). The corresponding theoretically calculated distributions are plotted as insets. A side view of the tightly focused beam propagating in $z$-direction (green arrow) is sketched in (f). Clockwise and counterclockwise transverse spinning of the local electric field vector left and right to the optical axis (gray arrows) yields an parallel $p_{s,E}^{z}$ (red arrow) on the optical axis.}
\end{figure}

\textit{Introduction.}\textemdash The angular momentum (AM) of light can be decomposed into two distinct parts, the spin and the orbital AM. While the spin AM is associated with circular polarization, i.e. with a spinning electric and magnetic field vector, the origin of orbital AM resides in the phase \mbox{gradient \cite{Belinfante1940,Bekshaev2007,Berry2009,Bekshaev2011}}. This distinction of the AM also implies the separation of the cycle-averaged linear momentum density of a monochromatic light field $\textbf{p}\propto\operatorname{Re}\left[\textbf{E}\times\textbf{H}^{*}\right]$, into spin and orbital parts, $\textbf{p}=\textbf{p}_{s}+\textbf{p}_{o}$. The orbital contribution $\textbf{p}_{o}$ equals the canonical momentum density, which is proportional to the local phase \mbox{gradient \cite{Bliokh2014}}. The solenodial spin part $\textbf{p}_{s}$ is referred to as Belinfante's spin momentum \mbox{density (BSMD) \cite{Belinfante1940}}. In beams that can be described within the paraxial approximation (here defined by possessing negligible longitudinal field components), BSMD is purely \mbox{transverse \cite{Bekshaev2007,Bliokh2014}}. This is plausible, since BSMD is defined by the rotation of the spin density, a quantity describing the local circular polarization of the light field, which in the paraxial regime is either parallel or anti-parallel to the beam's propagation direction.\\ 
However, these properties do not hold true in non-paraxial light fields or evanescent \mbox{waves \cite{Berry2009,Bliokh2010a}}. Actually, fields with a spinning axis transverse to the propagation direction (corresponding to transverse components of the spin density) have recently been investigated in the case of tightly focused polarization tailored vector \mbox{beams \cite{Winnerl2012,Banzer2013,Neugebauer2014a,Neugebauer2014}}, evanescent \mbox{waves \cite{Bliokh2014}}, surface plasmon \mbox{polaritons \cite{Bliokh2012}}, and whispering-gallery-mode \mbox{resonators \cite{Junge2013}}. In addition, they have proven to be a useful tool in nanooptics for controlling directional emission and waveguide coupling of dipole-like plasmonic antennas \cite{Rodriguez-Fortuno2013,Neugebauer2014,Petersen2014}.\\		
Besides these applications, the transverse spin density may also yield a longitudinal component of \mbox{BSMD \cite{Bliokh2014,Bliokh2012}}. While BSMD has long been considered a 'virtual' field quantity, since it does not exert force on an absorbing Rayleigh particle \mbox{(radius $\ll$ wavelength) \cite{Chaumet2000,Berry2009,Bliokh2014}}, recent theoretical and experimental studies indicate the possibility to measure BSMD directly by light-particle momentum transfer in the Mie-scattering regime \mbox{(radius $\approx$ wavelength) \cite{Bekshaev2013,Bliokh2014,Nieto-Vesperinas2010, Angelsky2012}}. Consequently, for the investigation of this intriguing longitudinal component of BSMD in an optical tweezers setup, and for the implementation of optical transverse spin dependent experiments \cite{Rodriguez-Herrera2010,Rodriguez-Fortuno2013,Neugebauer2014}, measurements of individual components of the spin density and BSMD in complex field distributions are highly desirable.\\
Here, we present direct measurements of the distributions of the transverse spin density components of tightly focused linearly and radially polarized monochromatic beams of light. In addition and, to the best of our knowledge, for the first time we determine the corresponding longitudinal component of BSMD from experimental data.\\ 
\textit{Theoretical foundation.}\textemdash To describe the distributions of the transverse spin density as well as BSMD for tightly focused fields, we start with the general definition of BSMD given by \cite{Belinfante1940,Bekshaev2007,Berry2009}
\begin{align}\label{eqn_ps}
\textbf{p}_{s}&\propto \nabla\times\operatorname{Im}\left[\epsilon_{0}\left(\textbf{E}^{*}\times\textbf{E}\right)+\mu_{0}\left(\textbf{H}^{*}\times\textbf{H}\right)\right]\text{.}
\end{align}
Equation \eqref{eqn_ps} depends on both the electric and magnetic field \cite{Berry2009}. Here, we only consider the electric contribution, \mbox{$\textbf{p}_{s,E}\propto \nabla\times\operatorname{Im}\left[\textbf{E}^{*}\times\textbf{E}\right]$}, since the plasmonic field probe used in the experiment described below predominantly reacts to the local electric field \cite{Lee2007}. The longitudinal component of $\textbf{p}_{s,E}$, here defined as $z$-component, is according to Eq. \eqref{eqn_ps} generated by transversely spinning fields and can be written as
\begin{align}\label{eqn_p}
p_{s,E}^{z}\propto \left(\partial_{x} s_{E}^{y}-\partial_{y} s_{E}^{x}\right)\text{,}
\end{align}
with the transverse components of the electric spin density defined by \cite{Bliokh2014}
\begin{subequations}\label{eqn_stokes}
\begin{align}
 s_{E}^{x}\propto & \operatorname{Im}\left[E_{y}^{*}E_{z}\right]\text{,}\\
 s_{E}^{y}\propto & \operatorname{Im}\left[E_{z}^{*}E_{x}\right]\text{.}
\end{align}
\end{subequations}
The upper index $i=x,y$ denotes the spinning axis of the electric field. Both parameters describing the transversely spinning fields are akin to the classical Stokes parameter, $S_{3}\propto s_{E}^{z}\propto\operatorname{Im}\left[E_{x}^{*}E_{y}\right]$, describing the longitudinal component of the spin density (circular polarization) in paraxial light beams \cite{Born1997,Bekshaev2007}. Actually, all three components of the spin density are proportional to the respective 3D Stokes parameters defined in \mbox{Ref. \cite{Setala2002,Roy2014,Sheppard2014}}. In the upcoming paragraph, we show how the distributions of $s_{E}^{x}$ and $s_{E}^{y}$ in the focal plane of a high numerical aperture focusing system can be determined via a nanoprobe scanning measurement, which also permits the calculation of $p_{s,E}^{z}$.\\
\textit{Experimental concept.}\textemdash Our experimental approach relies on the near-field interference of a transversely spinning dipole \cite{Rodriguez-Fortuno2013,Mueller2013,Neugebauer2014}. If the dipole is in close proximity to a second medium with a higher refractive index (here: glass, $n=1.5$), the evanescent fields can be partially converted into propagating waves observable in the far-field, in particular, in the region above the critical angle indicated by the red arcs in Fig. \ref{fig_setup}(a). The critical angle is outlined by the dashed black line and corresponds to a numerical aperture (NA) of 1. 
The interface between both dielectric media can be considered planar with the surface normal $\textbf{z}$ pointing into the glass half space. For the chosen parameters, the region above the critical angle is defined by $k_{\bot}/k_{0}=\text{NA}\in \left[1,1.5\right]$, with the transverse wavenumber $k_{\bot}=(k_{x}^2+k_{y}^{2})^{-1/2}$ and the wavenumber in vacuum $k_{0}=\omega/c_{0}$. The intensity emitted into the far-field consists of two polarization components, the transverse magnetic (TM) and transverse electric (TE) component:
\begin{align} \label{eqn_I}
I\left(k_{x},k_{y}\right)\propto \left|E_{\text{TM}}\right|^{2}+\left|E_{\text{TE}}\right|^{2}\text{.}
\end{align} 
Considering only one specific transverse wavenumber above the critical angle, $k_{\bot}=k_{\bot}'$, which implies the restriction to a ring around the optical axis of the system in $k$-space, we can obtain a simplified expression for the far-field intensity \cite{Neugebauer2014},
\begin{subequations}
\begin{align}\label{eqn_ETM}
\left|E_{\text{TM}}\right|^{2}&\propto\left|\frac{\imath \left|k_{z}\right|k_{x}}{|\textbf{k}|k_{\bot}'}q_{x}+ \frac{\imath \left|k_{z}\right|k_{y}}{|\textbf{k}|k_{\bot}'}q_{y} -\frac{k_{\bot}'}{|\textbf{k}|} q_{z}\right|^{2}\text{,}\\
\left|E_{\text{TE}}\right|^{2}&\propto\left|
-\frac{k_{y}}{k_{\bot}'}q_{x} +\frac{k_{x}}{k_{\bot}'}q_{y} \right|^{2}\label{eqn_ETE} 
\text{.}
\end{align}
\end{subequations}
The components of the electric dipole moment \mbox{$\textbf{q}=\left(q_{x},q_{y},q_{z}\right)$} have herein complex values, and the longitudinal component of the $k$-vector is defined by \mbox{$k_{z}=[(k_{0}n)^{2}-k_{\bot}'^{2}]^{1/2}$}.\\
As can be seen from \mbox{Eq. \eqref{eqn_ETE}}, the transverse electric component $\left|E_{\text{TE}}\right|^{2}$ is always symmetric with respect to the $z$-axis. In contrast, the transverse magnetic component $\left|E_{\text{TM}}\right|^{2}$ defined in \mbox{Eq. \eqref{eqn_ETM}} can be asymmetric (directional), because of the interference of the longitudinal dipole moment with the transverse dipole moments \cite{Rodriguez-Fortuno2013,Mueller2013,Neugebauer2014}. By comparing the intensity scattered into opposite directions along the $x$- and $y$-axis, \mbox{$\Delta I^{x}=I(0,-k_{\bot}')-I(0,k_{\bot}')$} and \mbox{$\Delta I^{y}=I(k_{\bot}',0)-I(-k_{\bot}',0)$}, we result in a simple measure for the directionality,
\begin{subequations}
\label{eqn_Idiff1}
\begin{align}
\label{eqn_Idiff1a}
\Delta I^{x} &\propto \operatorname{Im} \left[q_{y}^{*}q_{z}\right]\text{,}\\ \label{eqn_Idiff1b}
\Delta I^{y} &\propto \operatorname{Im} \left[q_{z}^{*}q_{x}\right]\text{.} 
\end{align}
\end{subequations}
The expressions for the directionality parameters in Eq. \eqref{eqn_Idiff1} are formally equivalent to the formulas for the transverse spin densities in Eq. \eqref{eqn_stokes}, the only difference being the dependence on the transversely spinning dipole moments, in lieu of the transversely spinning electric fields. For that reason we choose the same notation, with the upper indices $x$ and $y$ representing the respective spinning axis of the dipole. In our experiment, we utilize this equivalence between Eq. \eqref{eqn_Idiff1} and Eq. \eqref{eqn_stokes} to measure the spatial distribution of the transverse electric spin density and, accordingly, the longitudinal electric component of BSMD. For that purpose we use a sub-wavelength, dipole-like plasmonic particle as field probe, and choose the wavelength of the incoming beam to be close to the electric dipole resonance of the particle. In these conditions, it primarily gets excited by the local electric field. Its polarizability and, for this reason, its dipole moment is proportional to the local electric field vector, $\textbf{q}\propto \textbf{E}$ \cite{Lee2007}. With this approximation we can \mbox{rewrite Eqs. \eqref{eqn_Idiff1}} as:
\begin{subequations}
\label{eqn_Idiff2}
\begin{align}
\Delta I^{x} &\propto s_{E}^{x}\text{,} \\
\Delta I^{y} &\propto s_{E}^{y}\text{.}
\end{align}
\end{subequations}
By exploiting this simple proportionality law, we are able investigate the transverse spin density of the electric field in the focal plane of tightly focused radially and linearly polarized light beams. Both types of beams are often utilized in nanooptics \cite{Quabis2000,Banzer2010a,Bauer2013,Neugebauer2014}, and have interesting properties regarding $p_{s,E}^{z}$, as will be discussed below.\\ 
\textit{Experimental implementation.}\textemdash The experimental setup, similar to that described in our earlier studies \cite{Banzer2010a,Bauer2013}, is depicted in Fig. \ref{fig_setup}(b). The incoming paraxial beam (with either radial or linear polarization) is tightly focused by a first microscope objective with $\text{NA}=0.9$. In the focal plane, a spherical gold nanoparticle (radius 40 nm) sitting on a glass substrate is utilized to probe the transversely spinning electric fields. The wavelength of the beam, $\lambda=530$ nm, is close to the electric dipole resonance of the particle. The light scattered by the particle as well as the beam transmitted through the air-glass interface are collected and collimated by an oil immersion objective mounted from below. Since the second microscope objective has an NA of 1.3, the intensity distribution above the critical angle ($k_{\bot}/k_{0}=\text{NA}\in \left[1,1.3\right]$) can be imaged using an additional imaging lens and a camera (see exemplary image of the back focal plane in \mbox{Fig. \ref{fig_setup}(c))}.\\
While moving the particle through the focal plane ($xy$-plane), we measure the intensity distribution in the back focal plane for each particle position, restricting ourselves to the region above the critical angle as indicated. We average the detected intensity over four small separated regions in $k$-space (see white circles in Fig. \ref{fig_setup}(c)), yielding four intensity values ($I_{1}$, $I_{2}$, $I_{3}$, and $I_{4}$) corresponding to four different $k$-vectors. By calculating the difference signals $\Delta I^{x}=I_{3}-I_{1}$ and $\Delta I^{y}=I_{2}-I_{4}$, $s_{E}^{x}$ and $s_{E}^{y}$ are determined via Eqs. \eqref{eqn_Idiff2}.\\ 
\textit{Experimental and theoretical results.}\textemdash We use the aforementioned experimental scheme and reconstruction technique to investigate two different tightly focused beams under normal incidence. In theory, both beams have no longitudinally spinning fields, hence $s_{E}^{z}=0$ in the focal plane, and only the longitudinal component of BSMD, $p_{s,E}^{z}$, is different from zero.
\begin{figure}[htbp]
\centerline{\includegraphics[width=0.8\columnwidth]{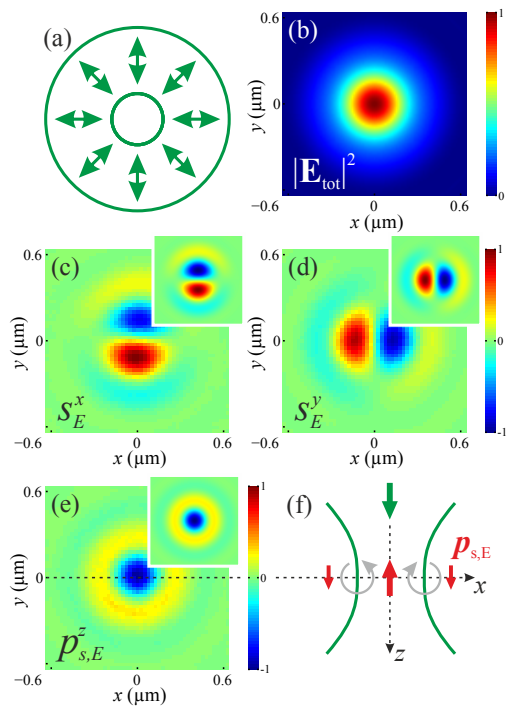}}
\caption{\label{fig_radial}(color). (a) The beam profile of the incoming paraxial radially polarized beam, and (b) the respective theoretically calculated focal energy density distribution of the electric field $\left|\textbf{E}_{\text{tot}}\right|^{2}$. The measured distributions of the two components of the transverse spin density $s_{E}^{x}$ and $s_{E}^{y}$ (normalized to their common maximum value) are depicted in (c) and (d), and the experimentally reconstructed component of Belinfante's spin momentum $p_{s,E}^{z}$ (normalized to its maximum value) is shown in (e). The corresponding theoretically calculated distributions are plotted as insets. A side view of the tightly focused beam propagating in $z$-direction (green arrow) is sketched \mbox{in (f)}. Counterclockwise and clockwise transverse spinning of the local electric field vector left and right to the optical axis (gray arrows) yields an anti-parallel $p_{s,E}^{z}$ (red arrow) on the optical axis.}
\end{figure}
Fig. \ref{fig_radial}(a) illustrates the polarization distribution of an incoming radially polarized beam before focusing, while Fig. \ref{fig_radial}(b) shows the numerically calculated electric field distribution $\left|\textbf{E}_{\text{tot}}\right|^{2}$ in the focus. Both distributions exhibit the same cylindrical symmetry with respect to the optical axis. However, a strong longitudinal field component appears in the center of the focal spot \cite{Quabis2000}. \mbox{In Fig. \ref{fig_radial}(c,d)} the measured transverse spin density components $s_{E}^{x}$ and $s_{E}^{y}$ are depicted. The experimental distributions exhibit a very good overlap with theory (see insets) demonstrating the high accuracy of the measurement approach. Using those results, the distribution of $p_{s,E}^{z}$ is reconstructed from the experimental data using Eq. \eqref{eqn_p} (see Fig. \ref{fig_radial}(e)). The comparison with the theoretical distribution (see inset) shows a very good overlap as well. An interesting result is that BSMD is negative in the center of the beam, which implies that, on the optical axis, BSMD is anti-parallel to the propagation direction of the radially polarized beam (see side view sketch in Fig. \ref{fig_radial}(f)). A similar effect can be seen in evanescent waves or surface plasmon polaritons \cite{Bliokh2014,Bliokh2012}. In theory, the magnetic field is zero on the optical axis of the tightly focused radially polarized beam, because of the azimuthal symmetry of the magnetic field vectors. Accordingly, the linear momentum density $\textbf{p}$ is zero as well. For this reason, the electric component of BSMD must have exactly the same value as the canonical momentum, but with the opposite sign, $p_{s,E}^{z}=-p_{o}^{z}$. The spin momentum and the orbital momentum are in equilibrium as indicated by the schematics in Fig. \ref{fig_radial}(f).\\
As an example for a case with significantly different distributions of $s_{E}^{x}$, $s_{E}^{y}$ and $p_{s,E}^{z}$, we consider a tightly focused linearly polarized beam. A sketch of the incoming paraxial beam is depicted in Fig. \ref{fig_linear}(a).
The calculated focal spot in Fig. \ref{fig_linear}(b) shows the typical elongated distribution caused by the breaking of the axial symmetry by the polarization state of the incoming beam \cite{Quabis2000}. Additionally, the distributions of $s_{E}^{x}$ and $s_{E}^{y}$ depicted \mbox{in Fig. \ref{fig_linear}(c,d)} are drastically changed in comparison to the radially polarized beam. The $x$-component of the electric spin density is very weak (it has nonzero values) and beyond the noise level of our experimental data. However, the measured distribution of $s_{E}^{y}$ has a very good overlap with the theoretical prediction. In contrast to the radially polarized beam \mbox{(see Fig. \ref{fig_radial}(d))}, the spinning directions of the electric field left and right to the optical axis have flipped their signs. As a result, the distribution of $p_{s,E}^{z}$ in Fig. \ref{fig_linear}(e) is positive on the optical axis, and therefore parallel to the canonical momentum \mbox{(see sketch in Fig. \ref{fig_linear}(f))}. This demonstrates that the sign of BSMD on the optical axis is linked to the symmetry and polarization distribution of the incoming beam.\\ 
\textit{Discussion and conclusions}\textemdash In this letter, we have measured the spatial distribution of the transverse components of the electric spin density in tightly focused vector beams. The high resolution of the nanoprobe scanning measurement permitted us the subsequent reconstruction of the longitudinal electric component of Belinfante's spin momentum density. By comparing radially and linearly polarized beams, both commonly used in nanooptics and optical tweezers experiments, we have shown that the transverse spin density and, consequently, BSMD can be tuned by the mode and polarization state of the incoming beam.\\ 
In a recent article, a full reconstruction technique of focal fields in terms of the amplitude and phase distributions of the electric field components has been \mbox{introduced \cite{Bauer2013}}. In principle, this scheme also enables the calculation of all components (electric and magnetic) of the spin density and the corresponding linear momentum \mbox{density \cite{Belinfante1940,Bekshaev2007,Berry2009}}. Nevertheless, it requires solving a complex set of equations from which the parameters can be deduced. Conversely, in this work we have shown that by restricting ourselves to the experimental reconstruction of the transverse component of the electric spin density, taking advantage of directional scattering, a simple difference signal measurement in the far-field is sufficient for an accurate high resolution determination of the longitudinal electric component of BSMD. In addition, the measurement technique can also be used to investigate spatially more extended field distributions.\\ 
In general, transversely spinning fields might be useful for achieving 3D control for the excitation of quantum dots or single atoms \cite{Junge2013}. Furthermore, they might find application in optical tweezers experiments. Especially, with the goal to determine individual forces and torques of an optical particle trap, a map of individual angular momentum components is crucial \cite{Angelsky2012,Zhao2007}.
\bibliography{bib}
\end{document}